\documentclass{PoS}
\newcommand{\MS}{\overline{\mathrm{MS}}}

\newcommand{\be}{\begin{equation}}
\newcommand{\ee}{\end{equation}}
\def\al{\alpha}

\def\bea{\begin{eqnarray}}
\def\eea{\end{eqnarray}}

\def\simg{{\ \lower-1.2pt\vbox{\hbox{\rlap{$>$}\lower6pt\vbox{\hbox{$\sim$}}}}\ }}
\def\siml{{\ \lower-1.2pt\vbox{\hbox{\rlap{$<$}\lower6pt\vbox{\hbox{$\sim$}}}}\ }} 

\title{The static quark self-energy at ${\cal O}(\al^{20})$ in perturbation theory}

\ShortTitle{The static quark self-energy at ${\cal O}(\al^{20})$ in perturbation theory}

\author{Gunnar S.\ Bali, Clemens~Bauer\\
Institut f\"ur Theoretische Physik, Universit\"at Regensburg\\
D-93040 Regensburg, Germany\\
        E-mail: \email{gunnar.bali@ur.de}\\
        \hspace*{1.2cm}\email{clemens.bauer@physik.uni-regensburg.de}}

\author{\speaker{Antonio Pineda}\\
Grup de F\'{\i}sica Te\`orica, Universitat
Aut\`onoma de Barcelona,\\E-08193 Bellaterra, Barcelona, Spain\\
E-mail: \email{AntonioMiguel.Pineda@uab.es}}

\abstract{
In Refs.~\cite{Bauer:2011ws,Bali:2013pla} we determined the infinite volume coefficients of the perturbative expansions
of the self-energies of static sources in the fundamental and 
adjoint representations in $\mathrm{SU}(3)$
gluodynamics to order $\al^{20}$. We used numerical stochastic perturbation theory \cite{DRMMO94}, where
we employed a new second order integrator and twisted boundary
conditions. The expansions were obtained in lattice regularization with the
Wilson action and two different discretizations of the 
covariant time derivative within the Polyakov loop. 
Overall, we obtained four different perturbative series.
For all of them the high order coefficients displayed the factorial growth 
predicted by the conjectured renormalon picture,
based on the operator product expansion. This enabled us
to determine the normalization 
constants of the leading infrared renormalons of heavy quark
and heavy gluino pole masses. Here we present improved determinations of the 
normalization constants and the perturbative coefficients by 
incorporating the four-loop $\beta$-function coefficient (which we also determine) in the fit function.}

\FullConference{31st International Symposium on Lattice Field Theory LATTICE 2013\\
		 July 29 -- August 3, 2013\\
		 Mainz, Germany}

\begin{document}


In Refs.~\cite{Bauer:2011ws,Bali:2013pla}, the existence of renormalons ($c_n \sim n!$) in 
quantum gluodynamics has been unambiguously established. 
The quantities studied were $\delta m$ and $\delta m_{\tilde g}$, the self-energies of static sources in the
fundamental ($R=3$) and adjoint ($R=8$) representations. Using Numerical Stochastic Perturbation Theory 
\cite{DRMMO94}, they were computed
up to ${\cal O}(\al^{20})$ and extrapolated to infinite volume
in the Wilson action lattice scheme:
\be
\delta m=\frac{1}{a}\sum_{n= 0}^{19}c^{(3,\rho)}_n\alpha^{n+1}(1/a)\,\mathrm{(fundamental)}, \qquad 
\delta m_{\tilde g}=\frac{1}{a}\sum_{n= 0}^{19}c^{(8,\rho)}_n\alpha^{n+1}(1/a)\,\mathrm{(adjoint)}\,,
\ee
where $a$ is the lattice spacing. $\rho=0$ and $\rho=1/6$ stand
for un-smeared and smeared temporal links, respectively,
within the Polyakov line,
\begin{equation}
\label{eq:defpol}
L^{(R)}(N_S,N_T)=\frac{1}{N_S^3}\sum_{\mathbf n}\frac{1}{d_R}
\mathrm{tr}\left[\prod_{n_4=0}^{N_T-1}U^R_4(n)\right]\,,
\end{equation}
used to determine $\delta m$ through the relation
\begin{equation}
\label{eq:deltam}
\delta m=-\lim_{N_S,N_T\rightarrow\infty}\frac{\ln\langle L^{(3,\rho)}(N_S,N_T)\rangle}{a N_T}\,,
\qquad  
\delta m_{\tilde g}=-\lim_{N_S,N_T\rightarrow\infty}\frac{\ln\langle L^{(8,\rho)}(N_S,N_T)\rangle}{a N_T}\ .
\ee
Renormalon dominance predicts that the large $n$ dependence of $c^{(R,\rho)}_n$ should be (see Ref.~\cite{Bali:2013pla} for notation and definitions)
\be\label{generalm2}
c^{(3/8,\rho)}_n \stackrel{n\rightarrow\infty}{=} N_{m/m_{\tilde g}}\,\left(\frac{\beta_0}{2\pi}\right)^n
\,\frac{\Gamma(n+1+b)}{
\Gamma(1+b)}
\left(
1+\frac{b}{(n+b)}s_1+\frac{b(b-1)}{(n+b)(n+b-1)}s_2+ \cdots
\right).
\ee
One of the major results of this analysis was the confirmation of this behavior and the 
determination of the normalization of the renormalon of the quark and gluelump 
($N_{m_{\tilde g}}=-N_{\Lambda}$) pole mass:
\bea
N^{\mathrm{latt}}_{m} &=& 19.0\pm 1.6\, , \quad (C_F/C_A)\, N^{\mathrm{latt}}_{\Lambda}
= -18.7\pm 1.8\, ,
\\
N^{\MS}_{m}&=& 0.660\pm 0.056\,, \quad (C_F/C_A)\, N^{\MS}_{\Lambda}=-0.649\pm 0.062\,.
\eea
These numbers are by more than ten standard deviations separated from zero,
consolidating, with this significance, the existence of the $d=1$ renormalon in gluodynamics for two different quantities. 
The above numbers are in agreement, within errors, with determinations
from continuum-like computations 
\cite{Pineda:2002se,Lee:2002sn,Bali:2003jq,Brambilla:2010pp}, but they have been obtained using completely
independent methods. 
This is highly
nontrivial given the factor $\simeq 29$ between the values of
$N_{m}$ and $N_{\Lambda}$ in both schemes. Moreover, in the $\MS$ scheme
the normalization was determined from the first few terms of the perturbative series only, while
in the lattice scheme $n\geq 9$ was required.  We remark that there
has always been some doubt about the reliability of determinations
of $N^{\MS}_{m}$ and $N^{\MS}_{\Lambda}$ from just very few orders of perturbation theory.
We have now provided an entirely independent
determination of these objects based on many orders of the expansion
that can systematically be improved upon. 
Moreover, for the first
time, it was possible to follow the factorial growth of the
coefficients over many orders, from around  $\al^{9}$ up to $\al^{20}$,
vastly increasing the credibility of the prediction.

We expect that
the renormalon dominance of perturbative expansions
sets in at much lower orders in the $\MS$ scheme
than in the lattice scheme. This is
supported by the consistency of our $N_m$-determination
with continuum estimates mentioned above.
Also the earlier onset of the asymptotics
in the $\MS$-like schemes devised in Ref.~\cite{Bali:2013pla}  is coherent with this assumption. In Ref.~\cite{Bali:2013pla} 
we turned this argument around to
estimate $\beta_3^{\mathrm{latt}}$ from the lattice-to-$\MS$ scheme conversion, 
assuming that $c_{3,\MS}$ was dominated by the renormalon
\be
\label{c3MSassump}
c_{3,\MS} \simeq 
N^{\MS}_{m}\,\left(\frac{\beta_0}{2\pi}\right)^3
\,\frac{\Gamma(4+b)}{
\Gamma(1+b)}
\left(
1+\frac{b}{(3+b)}s_1+ \frac{b(b-1)}{(3+b)(2+b)}s_2+\cdots
\right).
\ee
Using our central value $c_{3,\mathrm{latt}}^{(3,0)}=794.5$, we obtained\footnote{This number and $d_3=351$ 
correct Eq. (103) of the published version of Ref.~\cite{Bali:2013pla}.}
\be
\label{eq:beta3}
\beta_3^{\mathrm{latt}} \simeq -1.12 \times 10^6\,. 
\ee

\begin{figure}[t]
\centerline{\includegraphics[width=.8\textwidth,clip=]{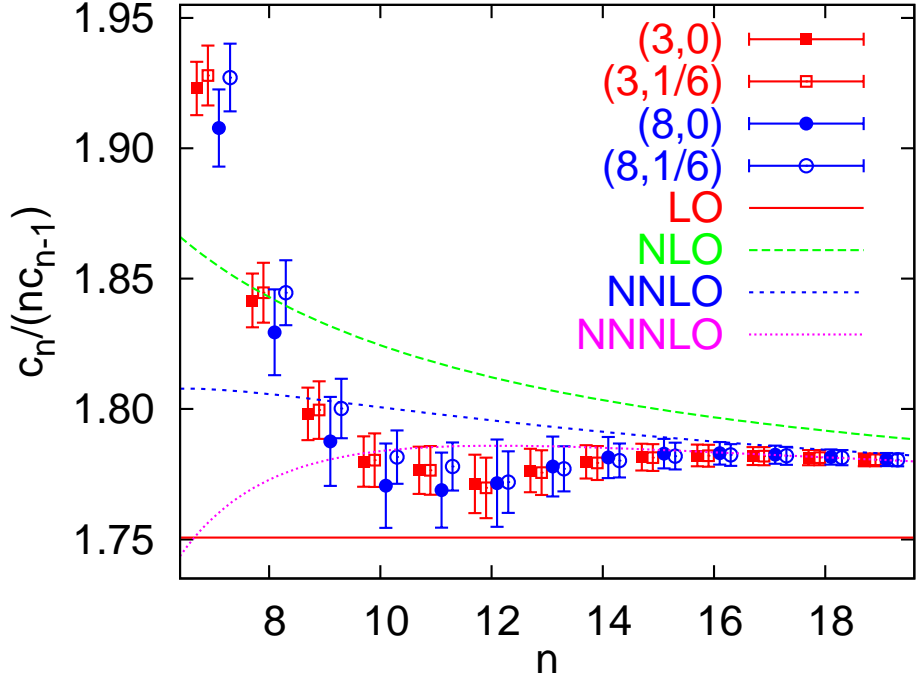}}
\caption{{\it The ratios
$c_n/(nc_{n-1})$ for the smeared and unsmeared, triplet and octet
fundamental static
self-energies, compared to the prediction
Eq.~(59) of Ref.~\cite{Bali:2013pla}  for the LO,
next-to-leading order
(NLO), NNLO and NNNLO of the
$1/n$ expansion.
For clarity, the data sets are slightly shifted horizontally.}
\label{n20}}
\end{figure}

\begin{figure}
\centerline{\includegraphics[width=0.8\textwidth,clip]{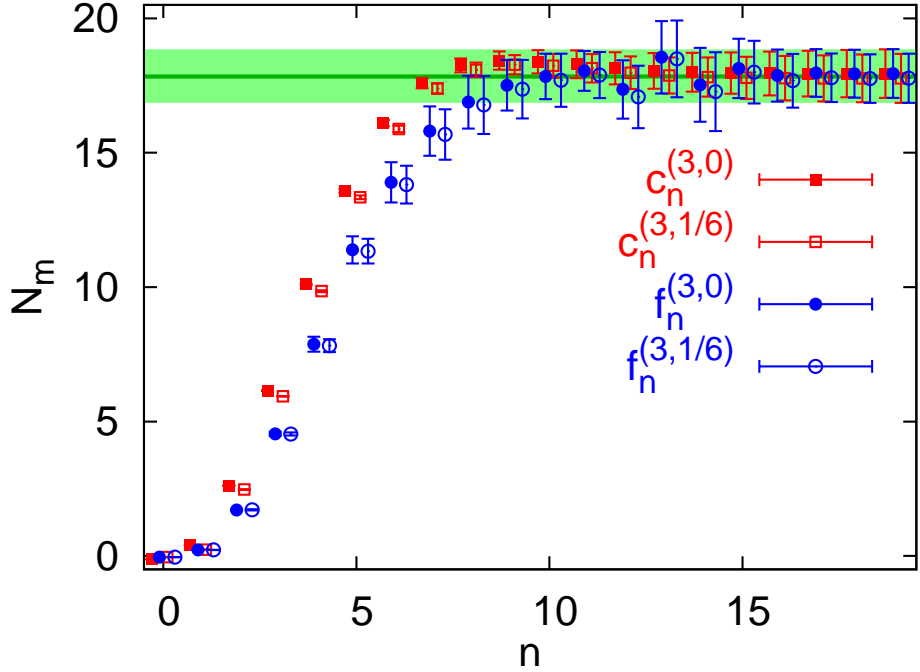}}
\caption{{\it $N_m$ in the lattice scheme, determined via Eq.~(4)
truncated at NNNLO from the
coefficients $c_n^{(3,0)},c_n^{(3,1/6)},f_n^{(3,0)}$ and $f_n^{(3,1/6)}$. The horizontal band is our final result quoted in Eq.~(10)
}
\label{fig:Nm}
}
\end{figure}

Crucial for the accurate determination of the coefficients $c_n^{(R,\rho)}$ (and 
the normalizations $N_{m,m_{\tilde g}}$) was the good theoretical control of the 
infinite volume extrapolation. Nevertheless, the final errors of the coefficients were 
still dominated by the systematics of this, due to 
the unknown higher order coefficients of the $\beta$ function:
in our fits we used the known values of $\beta_{0,1,2}$ and set $\beta_i=0$
from $\beta_3$ onwards.

We repeat the analysis of Ref.~\cite{Bali:2013pla} including the running due to 
$\beta_3^{\mathrm{latt}}$ in Eqs.~(68) and (70) of Ref.~\cite{Bali:2013pla}, and also its effect on the asymptotic analytic form of the renormalon in Eq. (\protect\ref{generalm2}). As $\beta_3^{\mathrm{latt}}$ is a free parameter, $N_m$ and the coefficients $c_n$ for $n\geq 3$ become functions of $\beta_3^{\mathrm{latt}}$. Selfconsistency with Eq. (\protect\ref{c3MSassump}) (assuming renormalon dominance at early orders in the $\MS$ scheme) fixes $\beta_3^{\mathrm{latt}}$ and we obtain  
$
\beta_3^{\mathrm{latt}} = -1.16 \times 10^6
$ (and $d_3=352$).
This value is almost identical to Eq. (\ref{eq:beta3}), illustrating the stability of the result to this procedure. Having gained this confidence, we will use this value to improve upon our analysis of Ref.~\cite{Bali:2013pla}. 

\begin{table}[h]
\caption{\it The infinite volume coefficients $c_n^{(R,\rho)}$, including
all systematic errors. The unsmeared $c_0$-values are fixed
using diagrammatic lattice perturbation theory.\label{tab:cnFinal}}
\begin{tabular}{c|cccc}
 &$c_n^{(3,0)}$&$c_n^{(3,1/6)}$&$c_n^{(8,0)}C_F/C_A$&$c_n^{(8,1/6)}C_F/C_A$\\\hline
$c_0$                   & 2.117274357  & 0.72181(99)  & 2.117274357  & 0.72181(99)\\
$c_{1}$         	    & 11.136(11)      & 6.385(10)      & 11.140(12)     & 6.387(10) \\
$c_{2}/10$           & 8.610(13)        & 8.124(12)      & 8.587(14)       & 8.129(12) \\
$c_{3}/10^2$       & 7.945(15)        & 7.671(11)      & 7.917(19)       & 7.682(13) \\
$c_{4}/10^3$       & 8.208(30)        & 8.009(28)      & 8.191(39)       & 8.010(32) \\
$c_{5}/10^4$       & 9.299(49)        & 9.135(49)      & 9.273(71)       & 9.116(55) \\
$c_{6}/10^6$       & 1.1478(86)      & 1.1324(87)    & 1.139(12)       & 1.1287(97) \\
$c_{7}/10^7$       & 1.545(16)        & 1.528(17)      & 1.521(21)       & 1.523(18) \\
$c_{8}/10^8$       & 2.276(32)        & 2.255(33)      & 2.225(42)       & 2.247(35) \\
$c_{9}/10^9$       & 3.684(68)        & 3.653(71)      & 3.580(90)       & 3.640(72)  \\
$c_{10}/10^{10}$& 6.56(15)          & 6.50(16)        & 6.34(20)         & 6.49(16)  \\
$c_{11}/10^{12}$& 1.281(36)        & 1.271(37)      & 1.234(45)       & 1.268(37) \\
$c_{12}/10^{13}$& 2.723(89)        & 2.699(91)      & 2.62(11)         & 2.697(92)  \\
$c_{13}/10^{14}$& 6.29(23)          & 6.23(24)       & 6.06(27)          & 6.23(24)  \\
$c_{14}/10^{16}$& 1.567(63)        & 1.552(64)     & 1.512(70)        & 1.553(64)  \\
$c_{15}/10^{17}$& 4.19(18)          & 4.15(18)       & 4.04(20)          & 4.15(18)  \\
$c_{16}/10^{19}$& 1.194(54)        & 1.182(55)     & 1.153(59)        & 1.184(55) \\
$c_{17}/10^{20}$& 3.62(17)          & 3.58(17)       & 3.49(18)          & 3.59(17)  \\
$c_{18}/10^{22}$& 1.160(57)        & 1.148(57)     & 1.121(61)        & 1.150(57)  \\
$c_{19}/10^{23}$& 3.92(20)          & 3.88(20)       & 3.79(21)          & 3.89(20) 
\end{tabular}
\end{table}

\begin{table}[h]
\caption{\it The $1/N_S$ correction coefficients $f_n^{(R,\rho)}$, including
all systematic errors. The unsmeared $f_0$-values are fixed
using diagrammatic lattice perturbation theory.\label{tab:fnfinal}}
\begin{tabular}{c|cccc}
 &$f_n^{(3,0)}$&$f_n^{(3,1/6)}$&$f_n^{(8,0)}C_F/C_A$&$f_n^{(8,1/6)}C_F/C_A$\\\hline
$f_0$                  & 0.7696256328  & 0.7811(69)  & 0.7696256328 & 0.7810(69)\\
$f_{1}$                & 6.075(78)         & 6.046(68)    & 6.124(87)        & 6.063(68) \\
$f_{2}/10$           & 5.628(91)         & 5.644(73)    & 5.60(11)          & 5.691(78) \\
$f_{3}/10^2$       & 5.867(99)         & 5.858(73)    & 6.00(17)          & 5.946(81) \\
$f_{4}/10^3$       & 6.40(23)           & 6.36(20)      & 6.65(40)          & 6.33(24) \\
$f_{5}/10^4$       & 7.79(35)           & 7.76(31)      & 7.73(67)          & 7.84(42) \\
$f_{6}/10^5$       & 9.91(53)           & 9.85(50)      & 9.73(99)          & 9.85(69) \\
$f_{7}/10^7$       & 1.389(81)         & 1.378(82)    & 1.35(15)          & 1.38(11) \\
$f_{8}/10^8$       & 2.11(12)           & 2.09(13)      & 2.05(22)          & 2.09(17) \\
$f_{9}/10^9$       & 3.50(19)           & 3.47(22)      & 3.35(36)          & 3.47(26)  \\
$f_{10}/10^{10}$ & 6.36(30)          & 6.31(35)       & 6.10(65)         & 6.31(41)  \\
$f_{11}/10^{12}$ & 1.264(52)        & 1.253(60)     & 1.21(12)         & 1.253(65) \\
$f_{12}/10^{13}$ & 2.61(16)          & 2.56(17)       & 2.57(32)         & 2.58(18)  \\
$f_{13}/10^{14}$ & 6.47(47)          & 6.44(50)       & 6.13(86)         & 6.43(51)  \\
$f_{14}/10^{16}$ & 1.53(12)          & 1.50(13)       & 1.49(21)         & 1.51(13)  \\
$f_{15}/10^{17}$ &  4.23(26)         & 4.20(27)       & 4.07(42)         & 4.20(28)  \\
$f_{16}/10^{19}$ & 1.189(64)        & 1.176(66)     & 1.151(89)       & 1.178(67) \\
$f_{17}/10^{20}$ & 3.62(18)          & 3.59(18)       & 3.50(21)         & 3.59(18)  \\
$f_{18}/10^{22}$ & 1.159(58)        & 1.148(58)     & 1.120(64)       & 1.149(59)  \\
$f_{19}/10^{23}$ & 3.92(20)          & 3.89(20)       & 3.79(21)         & 3.89(20) 
\end{tabular}
\end{table}

In Table \ref{tab:cnFinal} we display the infinite volume coefficients $c_n^{(R,\rho)}$, including
all systematic errors. The unsmeared $c_0$-values are fixed
using diagrammatic lattice perturbation theory.  The central values are obtained as in Ref.~\cite{Bali:2013pla} but 
including the running due to $\beta_3^{\mathrm{latt}}$ into Eqs.~(68) and (70) of this reference. We will take the 
errors of this fit as statistical ($\sigma^2_{\mathrm{stat.}}$). 
The quoted errors in table \ref{tab:cnFinal} have been computed as in Ref.~\cite{Bali:2013pla}. They result 
from summing statistical and theoretical uncertainties in quadrature.
Schematically, we have at each order $n$
\be
\label{error}
\sigma_{\mathrm{final}}=\sqrt{\sigma^2_{\mathrm{stat.}}+\sigma^2_{\beta}
+\sigma^2_{T}}
\,,
\ee
where $\sigma_{T}$ is the difference between central values of the fit with $\nu_T=11$ (our central value) and
$\nu_T=9$ (see Ref.~\cite{Bali:2013pla} for details). $\sigma_{\beta}$ is the difference between setting $\beta_3^{\mathrm{latt}}=0$ or not. We find
$\sigma_{\beta} \gg \sigma_{T}, \sigma_{\mathrm{stat.}}$, so 
that  the dominant error still stems from logarithmic
$N_S^{-1}\ln^i(N_S)$-corrections, due to our lack of knowledge of
$\beta_4^{\mathrm{latt}}$ etc.. 

The same analysis yields the $1/N_S$ correction coefficients $f_n^{(R,\rho)}$, where we determine the errors in the same way as for $c_n^{(R,\rho)}$. We display these results in Table \ref{tab:fnfinal}. The renormalon picture 
predicts that $c_n \simeq f_n$ for large $n$. This equality is achieved with a high degree of accuracy from $n=9$ onwards in all four cases (compare Tables \ref{tab:cnFinal} and \ref{tab:fnfinal}).

\begin{table}[h]
\caption{\it The infinite volume ratios $c_n^{(R,\rho)}/\left(nc_{n-1}^{(R,\rho)}\right)$, including all systematic errors. Note that $\beta_0/(2\pi)\approx 1.7507$.\label{tab:cnratioFinal}}
\begin{tabular}{c|cccc}
$n$&$c_n^{(3,0)}/\left(nc_{n-1}^{(3,0)}\right)$&$c_n^{(3,1/6)}/\left(nc_{n-1}^{(3,1/6)}\right)$&$c_n^{(8,0)}/\left(nc_{n-1}^{(8,0)}\right)$&$c_n^{(8,1/6)}/\left(nc_{n-1}^{(8,1/6)}\right)$\\
\hline
1   & 5.2594(53)     & 8.846(18)    & 5.2616(56)   & 8.848(18) \\
2   & 3.8662(61)     & 6.361(12)    & 3.8539(65)   & 6.364(12)  \\
3   & 3.0756(55)     & 3.1474(47)  & 3.0735(75)   & 3.1501(53) \\
4   & 2.5827(89)     & 2.6104(89)  & 2.586(12)     & 2.6067(99) \\
5   & 2.2659(95)     & 2.2812(98)  & 2.264(15)     & 2.276(12) \\
6   & 2.057(10)       & 2.066(11)    & 2.046(15)     & 2.064(13) \\
7   & 1.923(10)       & 1.928(11)    & 1.908(15)     & 1.927(13) \\
8   & 1.842(10)       & 1.845(11)    & 1.829(16)     & 1.845(12) \\
9   & 1.798(10)       & 1.780(11)    & 1.788(17)     & 1.800(11)  \\
10 & 1.7798(97)     & 1.780(10)    & 1.771(16)     & 1.782(10) \\
11 & 1.7765(91)     & 1.7765(94)  & 1.769(14)     & 1.7780(92)\\
12 & 1.771 (11)      & 1.770(12)    & 1.772(17)     & 1.772(12) \\
13 & 1.7764(83)     & 1.7756(86)  & 1.778(11)     & 1.7770(86)\\
14 & 1.7797(64)     & 1.7793(65)  & 1.7814(79)   & 1.7802(65)\\
15 & 1.7816(51)     & 1.7814(52)  & 1.7829(57)   & 1.7819(51)\\
16 & 1.7822(42)     & 1.7821(42)  & 1.7830(43)   & 1.7824(42)\\
17 & 1.7820(35)     & 1.7819(35)  & 1.7825(35)   & 1.7821(35)\\
18 & 1.7813(29)     & 1.7813(29)  & 1.7816(29)   & 1.7814(29)\\
19 & 1.7805(25)     & 1.7805(25)  & 1.7806(25)   & 1.7805(25)
\end{tabular}
\end{table}

In Table \ref{tab:cnratioFinal} we display the infinite volume $c_n^{(R,\rho)}/(nc_{n-1}^{(R,\rho)})$-ratios. 
The central values are trivially deduced from Table \ref{tab:cnFinal}. The statistical errors are obtained from the global fit to the
volume dependence, and include the statistical correlations between the different $n$-value $c_n^{(R,\rho)}$ coefficients. The total error is obtained as 
before, using Eq. (\ref{error}). In Fig.~\ref{n20} we display these ratios and compare them with the renormalon expectations (Eq. (59) of Ref.~\cite{Bali:2013pla}). We see that the agreement is fantastic. This 
means that, for large $n$, the coefficients are very well approximated by Eq.~(\ref{generalm2}), which we can use to fix $N_{m,\Lambda}(n)$. For large $n$ the result should be independent of $n$. We confirm this behavior in Fig. \ref{fig:Nm}. Working as in 
Ref.~\cite{Bali:2013pla} we obtain accurate determinations of the normalization constants of the renormalon. 
They read
\bea
\label{Nmfinal}
N^{\mathrm{latt}}_{m} &= 17.9\pm 1.0\, , \quad (C_F/C_A)\, N^{\mathrm{latt}}_{\Lambda}
= -17.6\pm 1.2\, ,\\
N^{\MS}_{m}&= 0.620\pm 0.035\,, \quad (C_F/C_A)\, N^{\MS}_{\Lambda}=-0.610\pm 0.041\,.
\eea
We stress that the $N_m$-value is by 18 standard
deviations different from zero!
Other combinations of interest are (see Eqs.~(56) and  (58) of Ref.~\cite{Bali:2013pla})
\be
N^{\MS}_{V_s}= -1.240\pm 0.069 \,, \quad N^{\MS}_{V_o}=0.13\pm 0.12
\,.
\end{equation}
The errors of the coefficients, ratios and $N_{m,\Lambda}$ are still
dominated by the systematics, though now they are reduced, relative
to Ref.~\cite{Bali:2013pla}.
Our previous central values (setting $\beta_3^{\mathrm{latt}}=0$)
agree within one standard deviation
with the new, improved numbers above.

We are now in the position to predict the four-loop relation between the pole and the $\MS$ mass in the limit of zero flavours, $r_3/m_{\MS}$ (see Eq. (43) of Ref.~\cite{Bali:2013pla} for notation) using Eq. (\ref{c3MSassump}). We obtain
\be
c_3^{\MS}=r_3/m_{\MS}=37.9(2.2)
\,,
\ee
where the error is dominated by the uncertainty of $N_m$ (the effect due to $1/n$ effects in the asympotic formula is subleading). 
This number is in perfect agreement with (-1/2 times) the number quoted in Eq. (4) of Ref. \cite{Pineda:2002se}. Once we have this value for $c_3^{\MS}$ we can determine $\beta_3^{\mathrm{latt}}$ as discussed around Eq. (\ref{c3MSassump}). We obtain\footnote{Note that the impact of the $\beta_3^{\mathrm{latt}}$-error
on the infinite volume coefficients
is clearly negligible compared with $\sigma_{\beta}$, the difference between the evaluations setting $\beta_3^{\mathrm{latt}}=0$ or not.} ($d_3=352(3)$)

\be
\label{b3final}
\beta_3^{\mathrm{latt}}=-1.16(12)\times 10^6
\,.
\ee 
The error is (conservatively) determined by linearly adding the errors due to $N_m$, $c_2^{\rm latt}$ and $c_3^{\rm latt}$ 
(again the $1/n$ corrections are negligible in comparison), even though they are correlated. 

Eq. (\ref{b3final}) is consistent with 
the value 
$
\beta_3^{\mathrm{latt}}=-1.55(19)\times 10^6
$ 
obtained in Ref.~\cite{Guagnelli:2002ia}, which we had been unaware of
at the time we wrote Ref.~\cite{Bali:2013pla}.
This number was found from a non-perturbatively
determined step-scaling function which allowed to compute $\alpha(a^{-1})$ for inverse 
lattice spacings up to $a^{-1} \siml 50$ GeV. Note though that such a high value for $-\beta_3^{\mathrm{latt}}$ would be in tension with the renormalon dominance of $c_3^{\MS}$. Therefore, we cannot avoid to remark that smaller values for $-\beta_3^{\mathrm{latt}}$ are obtained in Ref.~\cite{Guagnelli:2002ia} by restricting the fit range to the points at smaller lattice spacings (but then less points are available). 

{\bf Acknowledgments}.
This work was supported by the German DFG
Grant SFB/TRR-55, the Spanish 
Grants FPA2010-16963 and FPA2011-25948, the Catalan Grant SGR2009-00894
and the EU ITN STRONGnet 238353.


\begin{thebibliography}{99}

\bibitem{Bauer:2011ws} 
  C.~Bauer, G.~S.~Bali and A.~Pineda,
  Phys.\ Rev.\ Lett.\  {\bf 108}, 242002 (2012)
  [arXiv:1111.3946 [hep-ph]].

\bibitem{Bali:2013pla} 
  G.~S.~Bali, C.~Bauer, A.~Pineda and C.~Torrero,
  Phys.\ Rev.\ D {\bf 87}, 094517 (2013)
  [arXiv:1303.3279 [hep-lat]].

\bibitem{DRMMO94}
F.~Di~Renzo, E.~Onofri, G.~Marchesini and P.~Marenzoni,
Nucl.\ Phys.\ B \textbf{426}, 675 (1994)
[arXiv:hep-lat/9405019].

\bibitem{Pineda:2002se}
  A.~Pineda,
  J.\ Phys.\ G \textbf{29}, 371 (2003)
  [arXiv:hep-ph/0208031].

\bibitem{Lee:2002sn} 
  T.~Lee,
  Phys.\ Rev.\ D \textbf{67}, 014020 (2003)
  [arXiv:hep-ph/0210032].

\bibitem{Bali:2003jq}
  G.~S.~Bali and A.~Pineda,
  Phys.\ Rev.\  D \textbf{69}, 094001 (2004)
  [arXiv:hep-ph/0310130].

\bibitem{Brambilla:2010pp} 
  N.~Brambilla, X.~Garcia i Tormo, J.~Soto and A.~Vairo,
  Phys.\ Rev.\ Lett.\  \textbf{105}, 212001 (2010)
  [Erratum-ibid.\  \textbf{108}, 269903 (2012)]
  [arXiv:1006.2066 [hep-ph]].

\bibitem{Guagnelli:2002ia} 
  M.~Guagnelli, R.~Petronzio and N.~Tantalo,
  Phys.\ Lett.\ B {\bf 548}, 58 (2002)
  [hep-lat/0209112].

\end{thebibliography}
\end{document}